\documentclass[prd,nofootinbib,showkeys,showpacs]{revtex4}
\usepackage{amsmath}
\usepackage{dsfont}               

\newcommand{\platzoben}{$\vphantom{{M^M}^M}$}
\newcommand{\mehrplatzoben}{$\vphantom{{{M^M}^M}^M}$}

\begin{document}

\title{Properties of the vector meson nonet at large $N_c$ beyond the chiral limit}
\author{Stefan Leupold}
\affiliation{Institut f\"ur Theoretische Physik, Universit\"at Giessen, Germany}

\begin{abstract}
Masses and especially coupling constants of the vector meson nonet are determined in the 
large-$N_c$ limit, but beyond the chiral limit taking into account terms up to
quadratic order in the Goldstone boson masses. With two input parameters five
coupling constants for hadronic and dilepton decays are determined which agree very
well with the experimental results. The obtained parameters are also used to calculate
the pion and kaon decay constant in the large-$N_c$ limit. 
A consistent picture is only obtained, if 
the correct assignment of the
$N_c$-dependence of the electromagnetic charges of the quarks is taken into account.
\end{abstract}
\pacs{11.30.Rd, 12.40.Vv, 11.15.Pg}
\keywords{Chiral symmetry, meson properties, vector-meson dominance,
large-Nc expansion, chiral perturbation theory}

\maketitle

\section{Introduction}

It is nowadays common wisdom that 
the spontaneous breaking of chiral symmetry leads to the appearance of a Goldstone
boson octet. This causes a 
large gap in the excitation spectrum of the observed hadrons. The lowest pseudoscalar 
octet appears to be light\footnote{The meson masses deviate from zero on account of 
the (small) current quark masses which explicitly break the chiral symmetry.}
while all other hadrons are heavy. Therefore at low energies QCD reduces to an 
effective theory where only the pseudoscalar mesons appear which interact with each 
other and with external sources. Spontaneous chiral symmetry breaking also demands that 
the meson self-interaction vanishes with vanishing energy. Therefore a systematic 
expansion in terms of the derivatives of the meson fields is possible. These 
considerations lead to the effective Lagrangian of chiral 
perturbation theory ($\chi$PT) \cite{gasleut1,gasleut2}
presented in the following for three light flavors:
\begin{equation}
  \label{eq:chipt}
{\cal L}_{\chi \rm PT} = {\cal L}_1 + {\cal L}_2 + \mbox{higher order derivatives}
\end{equation}
with the term which contains two derivatives of the Goldstone boson fields,
\begin{equation}
  \label{eq:chipt2}
{\cal L}_1 = {1\over 4} \,F_0^2 \,
{\rm tr} (\nabla_\mu U^\dagger \nabla^\mu U) + \ldots  \,,
\end{equation}
and all possible terms which contain four derivatives,
\begin{eqnarray}
{\cal L}_2 & = & L_1 [{\rm tr}(\nabla_\mu U^\dagger \nabla^\mu U)]^2 
+ L_2 \, {\rm tr}(\nabla_\mu U^\dagger \nabla_\nu U) 
      \, {\rm tr}(\nabla^\mu U^\dagger \nabla^\nu U) \nonumber \\
&& {}+ L_3 \, {\rm tr}(\nabla_\mu U^\dagger \nabla^\mu U
      \nabla_\nu U^\dagger \nabla^\nu U) 
+ \ldots  \,.
\label{eq:chipt4}
\end{eqnarray}
Note that we have only displayed explicitly the terms which are relevant for later use, 
i.e.~the ones
which remain present once all external fields and explicit chiral symmetry breaking terms
are put to zero. All the other terms are indicated by the dots in (\ref{eq:chipt2})
and (\ref{eq:chipt4}). In $U$ the pseudoscalar meson fields are encoded. 
$F_0$ denotes the pion decay constant in the chiral limit. 
We refer to \cite{gasleut2} for further details. In principle, the low-energy
constants $F_0$ and $L_1$-$L_3$ and the ones which we have not explicitly displayed
($B_0$ and $L_4$-$L_{10}$) can be obtained from QCD by integrating out all degrees of 
freedom besides the Goldstone
bosons. Such a task, however, would be more or less equivalent to solving QCD in the
low-energy regime. Lattice QCD \cite{Creutz:1984mg} has started to determine some of 
these low-energy constants (see e.g.~\cite{Giusti:2004yp} and references therein). 
In practice, one determines these coupling
constants from experiment \cite{gasleut1,gasleut2,Amoros:2001cp}
and/or from hadronic \cite{eckgas,Ecker:1989yg,Donoghue:1989ed,pelaez02} 
or quark models 
\cite{diakpet,espraf,Schuren:1992sc,Mueller:1994dh,Pich:1995bw,Peris:1998nj,Leupold:2003zb}.
Indeed, already in the seminal work \cite{gasleut2} it was pointed out that it is
difficult to pin down in particular all ten constants $L_1$-$L_{10}$ solely from
experimental inputs. Large-$N_c$ considerations were involved in addition yielding
\begin{equation}
  \label{eq:l1l2rel}
L_1 - {1 \over 2} \, L_2, L_4, L_6, B_0 = o(1) \,,
\end{equation}
while
\begin{equation}
  \label{eq:scall1fpi}
L_1, L_2, L_3, L_5, L_8, L_9, L_{10}, F_0^2 = O(N_c)  \,.
\end{equation}
Here $N_c$ denotes the number of colors.
We note in passing that $L_7$ is special since it involves the 
chiral anomaly. See \cite{Peris:1994dh,Kaiser:2000gs} for details.

At low energies $\chi$PT yields an excellent description of the hadron phenomenology
(see e.g.~the reviews \cite{Ecker:1995gg,Pich:1995bw,Scherer:2002tk}). 
However, as an expansion in powers of energies and momenta of the involved states
the approach is limited to the low-energy regime. One limitation comes from the 
resonances which
appear in the meson-meson scattering amplitudes. A derivative expansion like the
one present in the $\chi$PT Lagrangian (\ref{eq:chipt}) can only give 
polynomials in the kinematic variables while a resonance appears as a pole.
Therefore to make contact between the low-energy region governed by $\chi$PT and the 
region of mesonic resonances
non-perturbative methods are needed --- which, of course, introduce some 
model dependence. 

In the following, we will make use of the resonance saturation approach 
developed in \cite{eckgas,Ecker:1989yg} (cf.~also \cite{Donoghue:1989ed}). 
The purpose of the present work is to extend this approach beyond the chiral limit and
study the properties of the lowest-lying vector meson nonet ($\rho$, $\omega$, $K^*$ and
$\phi$). We will concentrate on the masses of the vector meson nonet and their 
hadronic and dilepton decays. Concerning the masses we basically repeat the 
calculations of \cite{Cirigliano:2003yq}. The new aspect of the present work are the
decay properties. The basic ideas are the following:
\begin{itemize}
\item A Lagrangian of vector mesons coupled to the Goldstone bosons is proposed.
Contact between this Lagrangian and $\chi$PT is made by integrating out the vector
mesons in the low-energy regime. 
This task can be performed systematically in the framework of a large-$N_c$ expansion. 
Following \cite{eckgas} we 
will work in leading order of $1/N_c$ throughout the present work. 
(In \cite{Rosell:2004mn}
the resonance saturation approach is extended beyond the large-$N_c$ limit, but
is restricted to the chiral limit and to two flavors.)
\item We perform all calculations beyond the chiral limit by including systematically 
terms which are of quadratic order in the Goldstone boson masses (linear order in the
current quark masses). Concerning the decay properties of the vector mesons this
constitutes the main new aspect of the present work. 
\item Additional high-energy constraints are used to relate various coupling 
constants \cite{Ecker:1989yg}. 
As we will see, a consistent picture 
(beyond the chiral limit) emerges only, if the correct $N_c$-dependence of the 
electromagnetic charges of the quarks 
\cite{Abbas:1990kd,Shrock:1995bp,Abbas:2000xk,Bar:2001qk} is taken into account.
This new finding (which apparently involves aspects of the electroweak model) does not
show up in the chiral limit.
\end{itemize}

The paper is organized in the following way: In the next section we will introduce
the basic ideas of resonance saturation by presenting the whole approach in the 
chiral limit. This section
can be regarded as a brief summary of the works \cite{eckgas,Ecker:1989yg} as far it
concerns our purposes. In section \ref{sec:vnonet-ressat} we extend the approach
beyond the chiral limit. Numerical results, especially for coupling constants, 
are presented in section \ref{sec:results}. Finally we summarize in 
section \ref{sec:sum}.

\section{Resonance saturation in the chiral limit}
\label{sec:ressat}

One way to generalize $\chi$PT to higher energies is to introduce additional degrees
of freedom, i.e.~mesonic resonances with various quantum numbers. Of course, they
have to be included in accordance with chiral symmetry.
Contact with $\chi$PT is made --- at least in principle --- by integrating out these
resonance fields. In that way, the low-energy coefficients of $\chi$PT are expressed
in terms of the resonance parameters \cite{eckgas,Donoghue:1989ed,Ecker:1989yg}. 
In practice, interacting fields cannot be 
integrated out exactly. Here the large-$N_c$ expansion comes into play:
Meson interaction vertices are more suppressed \cite{'tHooft:1974jz} the higher the 
number of external legs is. This leads to
the fact that resonance loops are subleading as compared to tree diagrams. In addition,
we recall that $\chi$PT is an expansion around the chiral limit in derivatives and 
masses of the Goldstone bosons. Both aspects together yield a systematic way to integrate
out resonances in the low-energy regime \cite{eckgas}.

From a phenomenological
point of view \cite{pdg04} one could introduce arbitrary many of such resonances.
On the other hand, it was already observed in \cite{gasleut1} for $SU(2)$-$\chi$PT
that the $\rho$-meson basically saturates all low-energy coefficients to which it 
contributes. This issue has been extended to $SU(3)$ and studied more systematically
in \cite{eckgas,Ecker:1989yg}. The lowest-lying scalar and vector mesons were included
in that analysis. The starting point is the following Lagrangian:
\begin{equation}
  \label{eq:chirlagrres}
{\cal L}_{\rm res.sat.} = {\cal L}_1 + {\cal L}_{\rm kin} + {\cal L}_{\rm int} \,.
\end{equation}
Here ${\cal L}_1$ is again the lowest-order Lagrangian of $\chi$PT as 
given in (\ref{eq:chipt2}) and ${\cal L}_{\rm kin}$ denotes the kinetic part of the
resonance Lagrangian (see \cite{eckgas} for details).\footnote{Note that already here
interactions between the resonances and the Goldstone bosons are included via a
chirally covariant derivative. It is, however, subleading in $1/N_c$ and not relevant
for our purposes.} The last term is given by
\begin{equation}
  \label{eq:resgoldint}
{\cal L}_{\rm int} = {1 \over 2 \sqrt{2}} 
\left( 
F_V {\rm tr}\left(V_{\mu\nu} f^{\mu\nu}_+ \right)
+ i G_V {\rm tr}\left(V_{\mu\nu} [u^\mu , u^\nu] \right)
\right) + \ldots
\end{equation}
where we have only displayed the contributions from vector mesons explicitly.
In the large-$N_c$ limit the lowest-lying vector mesons can be collected in a 
nonet. In the chiral limit this nonet is completely degenerate. 
Here this nonet (in the tensor representation) is encoded in
$V_{\mu\nu}$, the Goldstone bosons in $u_\mu$ and the external currents
in $f^{\mu\nu}_+$ (see \cite{eckgas} for details).

As shown in \cite{eckgas} the lowest-lying resonances practically saturate the
low-energy coefficients. E.g.~for $L_1$-$L_3$ one obtains (in leading order
of $1/N_c$)
\begin{equation}
  \label{eq:l12ressat}
L_2 = 2 L_1 = {G_V^2 \over 4 M_V^2}  
\end{equation}
and
\begin{equation}
  \label{eq:l3ressat}
L_3 = -{3 G_V^2 \over 4 M_V^2} + \mbox{contributions from scalar mesons} \,.
\end{equation}
Here $M_V$ denotes the mass of the vector meson nonet in the chiral limit.
Note that there are no contributions from scalar resonances to $L_2$.
In \cite{Ecker:1989yg} the question was addressed whether the 
results (\ref{eq:l12ressat}) and (\ref{eq:l3ressat}) depend on the details how the
resonances are introduced, e.g.~in which representation. It turned out that 
at least for vector mesons the results are model independent, if constraints
from high-energy QCD are involved in addition. Even more, further relations between
$F_V$, $G_V$ and $F_0$ can be obtained by studying the electromagnetic and the axial
form factor of the pion \cite{Ecker:1989yg}: In the framework of (\ref{eq:chirlagrres})
the former is given by 
\begin{equation}
  \label{eq:formpion}
F(t) = 1 + { F_V G_V \over F_0^2} {t \over M_V^2 -t}  \,.
\end{equation}
Assuming that $F(t)$ vanishes at infinity (a quite natural assumption for a form factor)
one gets
\begin{equation}
  \label{eq:fvgvrel}
F_V G_V = F_0^2  \,.
\end{equation}
The axial form factor describes the coupling of a Goldstone boson to a photon and
an axial-vector current and controls e.g.~the decay $\pi^+ \to e^+ \nu \gamma$.
For a pion in the chiral limit it is given by
\begin{equation}
  \label{eq:axform}
G_A(t) = \frac{F_V \, (2 G_V - F_V)}{M_V^2} + \mbox{contributions from axial vectors} \,.
\end{equation}
The vector meson contribution comes from the process where the photon transforms to
a vector meson (coupling constant $F_V$) which in turn couples to the pion and the
axial-vector current with strength $2 G_V - F_V$.
Again, we demand that $G_A$ vanishes at infinity. The additional contributions not 
displayed explicitly fulfill this requirement. Thus we obtain \cite{Ecker:1989yg}
\begin{equation}
  \label{eq:fvgvrel2}
F_V = 2 G_V
\end{equation}
which finally leads to
\begin{equation}
  \label{eq:fvgvfinal}
F_V = 2 G_V = \sqrt{2} F_0
\end{equation}
and therefore
\begin{equation}
  \label{eq:l12final}
L_2 = 2 L_1 = {F_0^2 \over 8 M_V^2}  \,.
\end{equation}

Since the tensor representation of vector mesons is rather unusual it is illuminating 
to translate the results into the language of a standard $\rho\pi\pi$- and 
$\rho\gamma$-Lagrangian \cite{klingl1}
\begin{equation}
  \label{eq:standlagr}
{\cal L}_{\rm int} = i g \rho^\mu \, 
\left(
\pi^+ \partial_\mu \pi^- - \pi^- \partial_\mu \pi^+ 
\right)
+ g^2 \, \rho^\mu \rho_\mu \pi^+ \pi^-
- {e \over 2g_\gamma} \, F_\rho^{\mu\nu} F^A_{\mu\nu}
\end{equation}
where $\rho^\mu$ denotes the neutral $\rho$-meson, $\pi^{\pm}$ the charged pions, 
$A_\mu$ the photon field,
$e$ the electromagnetic coupling
and $F_B^{\mu\nu}$ the field strength corresponding to the field $B_\mu = \rho_\mu$ or
$A_\mu$, respectively. 
The connections of $G_V$ and $F_V$ to the usual $\rho\pi\pi$ and $\rho \gamma$
couplings are provided by
\begin{equation}
  \label{eq:grhopipi}
g = {G_V M_V \over F_0^2}
\end{equation}
and
\begin{equation}
  \label{eq:fvgcon}
g_\gamma = {M_V \over F_V}  \,.
\end{equation}
Relations (\ref{eq:grhopipi}) and (\ref{eq:fvgcon}) can be obtained by calculating the 
decay widths $\Gamma(\rho \to \pi\pi)$ and $\Gamma(\rho \to e^+ e^-)$ 
in both approaches (\ref{eq:chirlagrres}) and (\ref{eq:standlagr}). 
The equations (\ref{eq:fvgvfinal}) and (\ref{eq:l12final}) translate to
\begin{equation}
  \label{eq:gggmvl2}
g^2 = g_\gamma^2 = {M_V^2 \over 2 F_0^2} = {1 \over 16 L_2} \,,
\end{equation}
i.e.~we have obtained the universality of the $\rho$-meson coupling and the 
KSFR relation (in the chiral limit) \cite{ks,fr}. The fact that
the $\rho$-meson couples (at least approximately) with the same strength to
pions and photons constitutes one important aspect of the vector meson dominance
picture (e.g.~\cite{herrmann,klingl1,Dung:1996rp}, 
see also comment in \cite{Ecker:1989yg}). We note in passing that the relations
(\ref{eq:gggmvl2}) have also been obtained recently in a more general framework
by demanding perturbative renormalizability in the sense of effective field theories 
of a Lagrangian involving pions, $\rho$-mesons, nucleons and 
photons \cite{Djukanovic:2004mm,Djukanovic:2005ag}.

\section{Extension to the vector meson nonet}
\label{sec:vnonet-ressat}

In the previous section we have determined vector meson properties in the combined
chiral and large-$N_c$ limit. 
In the present section we go beyond the chiral limit by systematically
including terms in linear order in the quark masses (quadratic order in the Goldstone 
boson masses). Still we restrict ourselves to the large-$N_c$ limit to keep things
manageable. In that framework the mass splitting of the lowest-lying vector meson nonet
(and of other multiplets) was studied in \cite{Cirigliano:2003yq} in the resonance
saturation approach. Here we go beyond
that work as we also include the splitting of the coupling
constants of the vector nonet. 

We shall first review the results from \cite{Cirigliano:2003yq} as far as they concern
the vector mesons: As already pointed out, in the large-$N_c$ limit the vector mesons 
can be collected in 
a nonet (for simplicity we drop Lorentz indices as long as possible):
\begin{equation}
  \label{eq:vn-nonet}
R = {1 \over \sqrt{3} } \, R_0 \mathds{1} + {1 \over \sqrt{2} } \, R_i \lambda_i  \,.
\end{equation}
In the chiral limit this nonet would be completely degenerate. Splitting 
effects in linear order in the quark masses can be systematically included. Concerning
the vector meson masses the free resonance Lagrangian is extended in the following
way:
\begin{equation}
  \label{eq:vn-freereslagr}
{\cal L}_{\rm free} = {1 \over 2} {\rm tr}(\nabla R \cdot \nabla R - M_V^2 \, R^2)
+ e_m^V \, {\rm tr}(\chi_+ R^2)  \,.
\end{equation}
For our purposes $\chi_+$ reduces to\footnote{We assume perfect isospin symmetry.}
\begin{equation}
  \label{eq:vn-chiplmass}
\chi_+ \to 4 B_0 {\cal M} = 4 B_0 \left(
  \begin{array}{ccc}
m_q & 0 & 0 \\
0 & m_q & 0 \\
0 & 0 & m_s
  \end{array}
\right)
\end{equation}
with \cite{gasleut1,gasleut2}
\begin{equation}
  \label{eq:vn-defb0}
B_0 = -{\langle \bar q q \rangle \over F_0^2}  \,.
\end{equation}
In the large-$N_c$ limit one obtains ideal mixing 
\begin{subequations}
    \label{eq:vn-idealmix}
  \begin{eqnarray}
R_\omega = R_{\rm non-strange} & = & {1 \over \sqrt{3}} \, (R_8 + \sqrt{2} R_0)  \\
R_\phi = - R_{\rm strange} & = & {1 \over \sqrt{3}} \, (\sqrt{2} R_8 - R_0)
  \end{eqnarray}
\end{subequations}
and the following mass splitting pattern for the vector meson masses:
\begin{subequations}
    \label{eq:vn-ressatmasses}
  \begin{eqnarray}
    \label{eq:vn-ressatmassrho}
M^2_\rho = M^2_\omega & = & M_V^2 - 4 e_m^V M_\pi^2   \,, \\
    \label{eq:vn-ressatmassphi}
M^2_\phi & = & M_V^2 - 4 e_m^V \, (2 M_K^2 - M_\pi^2 )  \,,  \\
    \label{eq:vn-ressatmasskstar}
M^2_{K^*} & = & M_V^2 - 4 e_m^V M_K^2   \,.
  \end{eqnarray}
\end{subequations}
For more details we refer to \cite{Cirigliano:2003yq}. Note that 
the $\rho$- and the $\omega$-meson are still degenerate. This degeneracy will also
hold for the coupling constants to which we will turn next. It is only lifted beyond
the large-$N_c$ limit \cite{Cirigliano:2003yq}.

So far we have only reviewed the results of \cite{Cirigliano:2003yq}. Now we extend
this approach by including also splitting terms for the couplings of the vector
mesons to Goldstone bosons and external currents, i.e.~we extend (\ref{eq:resgoldint})
to
\begin{eqnarray}
{\cal L}_{\rm int} & = & {1 \over 2 \sqrt{2}} 
\left( 
F_V \, {\rm tr}\left(V_{\mu\nu} f^{\mu\nu}_+ \right)
+ \frac12 \, d_F \, {\rm tr}\left(V_{\mu\nu} \{ f^{\mu\nu}_+ , \chi_+ \} \right) 
+ \frac12 \, f_F \, {\rm tr}\left(V_{\mu\nu}  [ f^{\mu\nu}_+ , \chi_+  ] \right) 
+ i \, G_V \, {\rm tr}\left(V_{\mu\nu} [u^\mu , u^\nu] \right)
\right. \nonumber \\
&& \hspace*{1cm} \left. {}
+  \frac{i}{2} \, d_G \, {\rm tr}\left(V_{\mu\nu} \{ \chi_+ , [u^\mu , u^\nu] \} \right)
+ \frac{i}{2} \, f_G \, {\rm tr}\left(V_{\mu\nu}  [ \chi_+ , [u^\mu , u^\nu]  ] \right)
+ 2 i \, e_G \, {\rm tr}\left(u^\mu V_{\mu\nu} \, u^\nu \chi_+ \right)
\right)  \,.
  \label{eq:vnon-resgoldint}
\end{eqnarray}
This leads to a plenty of different coupling constants for various processes.
In the following we will be concerned with the couplings of vector mesons to photons
and the hadronic decays of vector mesons. This leads to
\begin{subequations}
    \label{eq:vnon-fvspl}
  \begin{eqnarray}
    \label{eq:vnon-fvspl-rho}
F_{\rho\gamma} = F_{\omega\gamma} & = & F_V + 4 B_0 m_q d_F = F_V + 2 M_\pi^2 d_F \,, \\
    \label{eq:vnon-fvspl-phi}
F_{\phi\gamma} & = & F_V + 4 B_0 m_s d_F = F_V + 2 \, (2 M_K^2 - M_\pi^2) \, d_F 
  \end{eqnarray}
\end{subequations}
instead of the chiral limit value $F_V$ and 
\begin{subequations}
    \label{eq:vnon-gvspl}
  \begin{eqnarray}
    \label{eq:vnon-gvspl-rhopipi}
G_{\rho\pi\pi} & = & G_V + 4 B_0 m_q \,(d_G-e_G) = G_V + 2 M_\pi^2 \,(d_G-e_G) \,, \\
    \label{eq:vnon-gvspl-rhoKK}
G_{\rho KK} = G_{\omega KK} & = & G_V + 4 B_0 m_q d_G - 4 B_0 m_s e_G 
= G_V + 2 M_\pi^2 \,(d_G+e_G) - 4 M_K^2 e_G \,, \\
    \label{eq:vnon-gvspl-phiKK}
G_{\phi KK} & = & G_V + 4 B_0 m_s d_G - 4 B_0 m_q e_G = 
G_V + 4 M_K^2 d_G - 2 M_\pi^2 \,(d_G+e_G)  \,, 
  \end{eqnarray}
\end{subequations}
instead of $G_V$. Note that the explicit isospin factors, e.g.~the fact that 
$\rho$-mesons couple differently to kaons and pions are not included in the definitions
of the coupling constants. They have to be taken into account for the calculation of
the respective process of interest (see below).

In addition, we will study the coupling $A_{VG}$ of a neutral 
(photon-like) vector meson $V$ to a Goldstone boson $G$ and an axial-vector current.
Such coupling constants are given by
\begin{subequations}
    \label{eq:vnon-avspl}
  \begin{eqnarray}
    \label{eq:vnon-avspl-rhopi}
A_{\rho\pi} & = & F_V - 2 G_V + 4 B_0 m_q \,(d_F- 2 d_G + 2 e_G) = 
F_V - 2 G_V + 2 M_\pi^2 \,(d_F- 2 d_G + 2 e_G)  \,, \\
A_{\rho K} = A_{\omega K} & = & F_V - 2 G_V + 4 B_0 m_q \, (d_F - 2 d_G) + 8 B_0 m_s e_G 
\nonumber \\    \label{eq:vnon-avspl-rhoK}
& = & F_V - 2 G_V + 2 M_\pi^2 \, (d_F - 2 d_G - 2 e_G) + 8 M_K^2 e_G  \,, \\
A_{\phi K} & = & F_V - 2 G_V + 4 B_0 m_s \, (d_F - 2 d_G) + 8 B_0 m_q e_G 
\nonumber \\      \label{eq:vnon-avspl-phiK}
& = & F_V - 2 G_V + 4 M_K^2 \, (d_F - 2 d_G) - 2 M_\pi^2 \, (d_F - 2 d_G - 2 e_G)    \,.
  \end{eqnarray}
\end{subequations}
Above we have used the 
Gell-Mann--Oakes--Renner relations \cite{GOR,gasleut2}
\begin{subequations}
\label{eq:vn-GOR}
\begin{eqnarray}
- m_q \langle \bar q q \rangle & = & {1 \over 2} F_0^2 M_\pi^2 \,,
  \label{eq:vn-GORrho}
\\ 
- m_s \langle \bar q q \rangle 
& = & {1 \over 2} F_0^2 \, (2 M_K^2 - M_\pi^2) \,,
  \label{eq:vn-GORphi}
\\
- {1 \over 2} \, (m_s + m_q) \langle \bar q q \rangle 
& = & {1 \over 2} F_0^2 M_K^2   \,.
  \label{eq:vn-GORstar}
\end{eqnarray}
\end{subequations}
Note that consistent with our approximation we have used
\begin{equation}
  \label{eq:samecond}
\langle \bar q q \rangle := \langle \bar u u \rangle = \langle \bar d d \rangle 
\approx \langle \bar s s \rangle
\end{equation}
and
\begin{equation}
  \label{eq:mkmpimqms}
{M_K^2 \over M_\pi^2} \approx {m_s + m_q \over 2 m_q}  \,.
\end{equation}

Like for the chiral limit discussed above, we can utilize the high-energy behavior of
form factors to relate various coupling constants. We start our discussion with the 
pion. Its coupling to real and virtual photons is mediated by the $\rho$-meson only.
(In contrast, the kaons involve $\rho$, $\omega$ and $\phi$, see below.) 
We study first the electromagnetic form factor. Generalizing (\ref{eq:formpion})  it
is now given by
\begin{equation}
  \label{eq:formpion2}
F(t) = 1 + { F_{\rho\gamma} G_{\rho\pi\pi} \over F_\pi^2} {t \over M_\rho^2 -t}
\end{equation}
with the pion decay constant $F_\pi$. In the large-$N_c$ approximation the latter
is related to its chiral limit value $F_0$ by \cite{gasleut2}
\begin{equation}
  \label{eq:fpil5}
F_\pi^2 = F_0^2 + 8 L_5 M_\pi^2 + o(M_\pi^4)  \,.
\end{equation}
Demanding again that the form factor (\ref{eq:formpion2}) vanishes at high energies 
we obtain 
\begin{equation}
  \label{eq:peFFm}
d_F G_V + (d_G - e_G) F_V = 4 L_5
\end{equation}
in addition to the chiral limit relation (\ref{eq:fvgvrel}).

For the axial form factor of the pion the contribution of the $\rho$-meson
--- which is a constant in energy and should therefore vanish --- is proportional to
$A_{\rho\pi}$ as given in (\ref{eq:vnon-avspl-rhopi}). Thus we get
\begin{equation}
  \label{eq:paFFm}
d_F - 2 d_G + 2 e_G = 0 
\end{equation}
in addition to the chiral limit relation (\ref{eq:fvgvrel2}). Hence we can already
relate some of the coupling constants to the low-energy parameter $L_5$:
\begin{equation}
  \label{eq:resdfdgeg}
d_F = \frac{2 \sqrt2 L_5}{F_0} \,, \qquad d_G - e_G = \frac{\sqrt2 L_5}{F_0} \,.
\end{equation}

Next we turn to the kaons. Here the discussion becomes more subtle since 
$\rho$-, $\omega$- and $\phi$-mesons are involved. The demand for a proper 
high-energy behavior of the form factors constrains only the sum of the vector meson
contributions. Apparently the relative weight of the vector meson contributions
becomes an issue. Here the quark charges come into play and in particular their
$N_c$-dependence. We have to work out first
how the different flavor currents contribute to the electromagnetic current for an
arbitrary number of colors. We introduce currents which correspond to $\rho$, $\omega$
and $\phi$:\\
isovector current
\begin{equation}
  \label{eq:vnonet-currho}
j_\mu^\rho  := {1 \over 2} \, (\bar u \gamma_\mu u - \bar d \gamma_\mu d )  \,,
\end{equation}
non-strange isoscalar current
\begin{equation}
  \label{eq:vnonoet-curom}
j_\mu^\omega  := {1 \over 2} \, (\bar u \gamma_\mu u + \bar d \gamma_\mu d ) \,,
\end{equation}
strange isoscalar current
\begin{equation}
  \label{eq:vnonoet-curphi}
j_\mu^\phi  := {1 \over \sqrt{2}} \, \bar s \gamma_\mu s  \,.
\end{equation}
We decompose the quark charges into weak isospin and hypercharge. In the
standard model the quarks belong to the fundamental representation of weak isospin.
This yields
\begin{equation}
  \label{eq:vn-charges}
Q_u = Y_q + \frac12 \,, \qquad
Q_d = Q_s = Y_q - \frac12 \,.
\end{equation}
According to \cite{Abbas:1990kd,Shrock:1995bp,Abbas:2000xk,Bar:2001qk} 
the weak hypercharge of the quarks is given by
\begin{equation}
  \label{eq:weakhypch}
Y_q = \frac{1}{2 N_c}  \,.
\end{equation}
This assignment ensures the cancellation of anomalies in an electroweak theory with
an arbitrary number of quark colors. One purpose of the present work is to demonstrate
that one gets a similar (somewhat weaker) relation between hypercharge and number of
colors within the vector meson framework outlined here. Therefore we shall not make
use of (\ref{eq:weakhypch}) in the following.
The electromagnetic current can now be decomposed:
\begin{equation}
  \label{eq:vn-emcur}
j_\mu^{\rm el} = Q_u \, \bar u \gamma_\mu u + Q_d \, \bar d \gamma_\mu d +
Q_s \, \bar s \gamma_\mu s 
= (Q_u-Q_d) \, j_\mu^\rho + (Q_u + Q_d) \, j_\mu^\omega + \sqrt{2} \, Q_s \, j_\mu^\phi
= j_\mu^\rho + 2 Y_q \, j_\mu^\omega 
+ {1 \over \sqrt{2}} \, \left( 2 Y_q -1 \right) j_\mu^\phi \,.
\end{equation}
Demanding that the electromagnetic form factor of the charged kaon vanishes at
high energies leads to
\begin{equation}
  \label{eq:eFFKm}
F_{\rho\gamma} G_{\rho KK} + 2 Y_q F_{\omega\gamma} G_{\omega KK}
+ {1 \over \sqrt{2}} \, \left( 2 Y_q -1 \right) F_{\phi \gamma} \, 
  (-\sqrt{2}) G_{\phi KK} = 2 F_K^2
\end{equation}
with the kaon decay constant $F_K$. The factor $(-\sqrt{2})$ in front of $G_{\phi KK}$
accounts for the different coupling of the kaon pair to $\rho$ and $\phi$, respectively
(isospin). 
Note that relation (\ref{eq:eFFKm}) is only correct in leading order of the 
$1/N_c$-expansion. In this limit the kaon decay constant is given by 
(cf.~\eqref{eq:fpil5}) \cite{gasleut2}
\begin{equation}
  \label{eq:fKl5}
F_K^2 = F_0^2 + 8 L_5 M_K^2   \,.
\end{equation}
Using all knowledge obtained so far we get
\begin{equation}
  \label{eq:fzyq}
Y_q B_0 \, (m_s - m_q)(G_V d_F + F_V d_G + F_V e_G) = 0 \,.
\end{equation}
In particular, all terms which are not proportional to $Y_q$ have vanished on account
of the corresponding relation (\ref{eq:peFFm}) for the pion.
There are two possible solutions to equation (\ref{eq:fzyq}) --- if one disregards
$B_0 = 0$ and $m_s = m_q$. 
Either the terms in the brackets vanish, 
\begin{equation}
  \label{eq:fpos}
G_V d_F + F_V d_G + F_V e_G \stackrel{?}{=} 0  \,,
\end{equation}
or $Y_q$. Since we work in leading order of the $1/N_c$-expansion the latter
possibility implies
\begin{equation}
  \label{eq:weakhypch2}
Y_q \stackrel{?}{=} o(1/N_c)  \,.
\end{equation}
Obviously, this would be in agreement with (\ref{eq:weakhypch}), but a somewhat weaker 
statement. In the following, we will collect further evidence that (\ref{eq:weakhypch2})
is right and (\ref{eq:fpos}) wrong. 

It is tempting to discuss next the electromagnetic form factor of the {\em uncharged}
kaon. However, we refrain from involving it. The reason is the following: 
From the point of view of chiral perturbation theory all orders contribute to the
form factors for the charged Goldstone bosons whereas the leading order does not
contribute to the form factor of the uncharged kaon. The constraints which we derive
for the vector meson coupling constants emerge from a cancellation between the
leading and the next-to-leading order contribution.\footnote{Of course, the
constraints emerge for high energies, i.e.~outside the realm of strict chiral 
perturbation theory. This is accounted for by using the full propagator structure
of the vector mesons instead of a heavy vector meson mass expansion.} 
To derive further relations
from the electromagnetic form factor of the uncharged kaon one should involve also at 
least two orders in the chiral 
expansion which is beyond the scope of the present work. 

We turn to the axial form factor of the charged kaon. In the by now usual way we get
\begin{equation}
  \label{eq:achKFFm}
\frac{1}{M_\rho^2} \, F_{\rho\gamma} A_{\rho K} + 
\frac{1}{M_\omega^2} \, 2 Y_q F_{\omega\gamma} A_{\omega K} + 
\frac{1}{M_\phi^2} \, {1 \over \sqrt{2}} \, \left( 2 Y_q -1 \right) F_{\phi \gamma} \, 
  (-\sqrt{2}) A_{\phi K} = 0  \,.
\end{equation}
Using (\ref{eq:fvgvrel2}) and (\ref{eq:vnon-avspl}) we observe that all the
coupling constants $A_{...}$ are already of linear order in the current quark masses.
Therefore, the vector meson masses and the coupling constants $F_{...}$ in
(\ref{eq:achKFFm}) can be taken in the chiral limit. Finally we obtain
\begin{equation}
  \label{eq:yqse}
Y_q B_0 \, (m_s - m_q) (d_F - 2 d_G - 2 e_G) = 0  \,.
\end{equation}
Again, it is the corresponding condition (\ref{eq:paFFm}) for the pion which causes
all terms to vanish which are not proportional to $Y_q$. Again we conclude that
(\ref{eq:weakhypch2}) holds or 
\begin{equation}
  \label{eq:wrong2}
d_F - 2 d_G - 2 e_G \stackrel{?}{=} 0 \,.
\end{equation}
Suppose for a moment that (\ref{eq:weakhypch2}) does {\em not} hold. In this case,
we can use (\ref{eq:fvgvfinal}), (\ref{eq:resdfdgeg}), 
(\ref{eq:fpos}) and (\ref{eq:wrong2}) to obtain
\begin{equation}
  \label{eq:wrong3}
L_5 = d_F = d_G = e_G = 0 \qquad \mbox{(presumably wrong!)}
\end{equation}
From a principal point of view this possibility cannot be excluded. On the other hand,
there is up to now no QCD-motivated reason known why the low-energy constant $L_5$
should vanish. In fact, the experimental facts suggest that $L_5 \neq 0$ and has the
same order of magnitude as all other $L_i$'s 
given in (\ref{eq:scall1fpi}) \cite{gasleut2}. Thus we conclude that either 
(\ref{eq:fpos}) or (\ref{eq:wrong2}) is wrong (or both) which inevitably leads to
\begin{equation}
  \label{eq:weakhypch3}
Y_q = o(1/N_c)  \,.
\end{equation}

The situation can be summarized as following: 
In the chiral limit, (\ref{eq:eFFKm}) and (\ref{eq:achKFFm}) are always fulfilled, 
irrespective of the
$N_c$-dependence of the weak hypercharge. Beyond the chiral limit we could deduce
the scaling (\ref{eq:weakhypch3}) but not the exact relation (\ref{eq:weakhypch}).
It would be interesting to extend the approach presented here beyond the leading order
of $1/N_c$. It might then be possible to connect the weak hypercharge to hadronic
quantities, e.g.~to $M_V^2/F_0^2$. In this way the weak hypercharge, or to phrase it
differently: the number of colors, could be determined within a hadronic 
framework (cf.~the discussions in \cite{Bar:2001qk,Borasoy:2004ua,Borasoy:2004mf}).
Such an extension, however, is beyond the scope of the present work.

We still have one form factor to further constrain our coupling constants, namely
the axial form factor for the neutral kaons. We get
\begin{equation}
  \label{eq:aneKFFm}
\frac{1}{M_\rho^2} \, F_{\rho\gamma} A_{\rho K} + 
\frac{1}{M_\omega^2} \, 2 Y_q F_{\omega\gamma} \, (-1) A_{\omega K} + 
\frac{1}{M_\phi^2} \, {1 \over \sqrt{2}} \, \left( 2 Y_q -1 \right) F_{\phi \gamma} \, 
  \sqrt{2} A_{\phi K} = 0  \,.
\end{equation}
Note the sign changes relative to (\ref{eq:achKFFm}). This leads to 
\begin{equation}
  \label{eq:aneKFFm2}
(1-2 Y_q) (d_F - 2 d_G - 2 e_G) = 0  \,.
\end{equation}
We ignore the possibility $Y_q = 1/2$ and conclude
\begin{equation}
  \label{eq:aneKFFm3}
d_F - 2 d_G - 2 e_G = 0  \,.
\end{equation}
Together with (\ref{eq:resdfdgeg}) we get 
\begin{equation}
  \label{eq:resdfdgeg2}
d_F = \frac{2 \sqrt2 L_5}{F_0} \,, \qquad d_G = \frac{\sqrt2 L_5}{F_0} \,, 
\qquad e_G = 0  \,.
\end{equation}

To summarize we have obtained a consistent picture by
adopting the arguments of \cite{Ecker:1989yg} concerning the form factors and extending
them beyond the chiral limit. It was important to introduce the correct 
dependence (\ref{eq:weakhypch3}) of the hypercharge on $N_c$. Only in this way
the equations for the kaon form factors produce the same results as the ones for the 
pion form factors. 

Starting with five initially unknown constants $d_F$, $d_G$, $e_G$, $f_F$ and $f_G$ in
(\ref{eq:vnon-resgoldint}) we have managed to determine the first three of
them. It is actually not surprising that we got no constraints on the other two
constants: All form factors involved only the neutral vector mesons (due to their
coupling to photons). If the flavor matrix $V_{\mu\nu}$ has only diagonal entries,
it commutes with the mass matrix in (\ref{eq:vn-chiplmass}). Therefore, the
$f_F$- and $f_G$-terms in (\ref{eq:vnon-resgoldint}) vanish in such cases.
In principle, $f_G$ influences the hadronic width of the $K^*$, while $f_F$ appears in
the coupling of the $K^*$ to the quark current $\bar u \gamma_\mu s$ which in turn
is important e.g.~for the $\tau$-decay and for 
QCD sum rules \cite{shif79,Reinders:1984sr}. We postpone the determination of these
parameters to future work.

In the next section we will determine the coupling constants (\ref{eq:vnon-fvspl}) and
(\ref{eq:vnon-gvspl})
as far as they are easily accessible from experiment. We will confront the obtained
results with our relations (\ref{eq:resdfdgeg2}).

\section{Experimental results for vector meson masses and decays}
\label{sec:results}

In (\ref{eq:vn-ressatmasses}) the masses of the vector
mesons are expressed in terms of the two parameters $M_V$ and $e_m^V$. The former
is the vector meson mass in the chiral limit and the latter parameterizes the splitting
pattern. We fit these two parameters to the experimental values of the vector meson 
masses. The results are shown in table \ref{tab:vnonet}. Obviously a very good fit
is obtained with an average deviation of the squared masses from the fit of only
1.5\%. 

Again we stress that this is by no means a new result \cite{Cirigliano:2003yq}. 
The new aspects concern the decay properties of the vector mesons to which we turn now.
Indeed, in terms of only two parameters the relations (\ref{eq:resdfdgeg2}), 
(\ref{eq:vnon-fvspl}), (\ref{eq:vnon-gvspl}), (\ref{eq:fpil5}) and (\ref{eq:fKl5})
determine the hadronic and electromagnetic coupling constants of the neutral 
vector mesons and
the pion and kaon decay constant, in total --- as we will see --- seven quantities.
In the following, we will present the formulae for the vector meson decays. 
Subsequently we will fit our two free parameters to two of these decay widths and
predict the other ones.

The dilepton decay widths are given by
\begin{subequations}
    \label{eq:vn-dilepw}
  \begin{eqnarray}
    \label{eq:vn-dilepwrho}
\Gamma(\rho \to e^+ e^-) & = & 
{4 \pi \alpha^2 \over 3} \, {F_{\rho\gamma}^2 \over M_\rho} 
\,, \\    
    \label{eq:vn-dilepwom}
\Gamma(\omega \to e^+ e^-) & = & {4 \pi \alpha^2 \over 3} \, {1 \over 9} \, 
{F_{\omega\gamma}^2 \over M_\omega}  \,, \\    
    \label{eq:vn-dilepwphi}
\Gamma(\phi \to e^+ e^-) & = & {4 \pi \alpha^2 \over 3} \, {2 \over 9} \, 
{F_{\phi\gamma}^2 \over M_\phi}  \,.
  \end{eqnarray}
\end{subequations}
with the electromagnetic fine structure constant $\alpha \approx 1/137$. The different
factors in (\ref{eq:vn-dilepw}) are caused by the different contributions of the
currents (\ref{eq:vnonet-currho})-(\ref{eq:vnonoet-curphi}) to the electromagnetic
current \eqref{eq:vn-emcur}.
Recall that the weight factors are ${1 \over N_c} = {1 \over 3}$ for 
the $\omega$ and 
${1 \over \sqrt{2}} ( {1 \over N_c} -1 ) = - {\sqrt{2} \over 3}$ for the
$\phi$. 
Our strategy will be to calculate $F_{i\gamma}$ for $i=\rho,\omega,\phi$ 
in our approach 
and compare these values to the experimental ones obtained from \eqref{eq:vn-dilepw}.

Next we turn to the hadronic decay widths. We consider the decays of the
vector mesons into two Goldstone bosons. The $\omega$ decays dominantly into three
pions due to an anomalous coupling. We do not consider this effect here. 
The $\omega$-decay into two pions caused by $\rho$-$\omega$ mixing is beyond the leading
$1/N_c$ effects and therefore also not included.
We get
\begin{subequations}
    \label{eq:vn-hadrw}
  \begin{eqnarray}
    \label{eq:vn-hadrwrho}
\Gamma(\rho^0 \to \pi^+ \pi^-) & = & {1 \over 48 \pi} 
{G_{\rho\pi\pi}^2 M_\rho^2 \over F_\pi^4} \, M_\rho \, 
\left(1-{4 M_\pi^2 \over M_\rho^2} \right)^{3/2}  \,, \\
    \label{eq:vn-hadrwphi}
\Gamma(\phi \to K^+ K^-) & = & {1 \over 96 \pi} 
{G_{\phi KK}^2 M_\phi^2 \over F_K^4} \, M_\phi \, 
\left(1-{4 M_K^2 \over M_\phi^2} \right)^{3/2}  
\,.
  \end{eqnarray}
\end{subequations}
Again, we will calculate the coupling constants $G_{...}$ 
in our approach and compare the results to the experimental ones obtained from 
(\ref{eq:vn-hadrw}). 

Note that the hadronic decay widths determined above are $O(1/N_c)$, i.e.~suppressed, 
as it should be \cite{'tHooft:1974jz}. This
can be most easily seen by recalling that
the vector meson masses are $O(1)$ and the pion and kaon decay constants $O(\sqrt{N_c})$.
Suppressed decay widths mean that the vector mesons are stable in the large-$N_c$ 
limit \cite{'tHooft:1974jz}. In turn, this indicates that our large-$N_c$ description
should work best for narrow states and less accurate for broad resonances. Therefore
we determine our free parameters from fits to partial decay widths of the most narrow
states.
\begin{table}[thb]
\centering
\begin{tabular}{|c|c|c|c|}
\hline
quantity & theory & experiment & ref. \\ \hline \hline
$M_V$ [GeV] \platzoben  & 0.766  & --  &   \\ \hline
$e_m^V$  \platzoben  & $-0.233$ & --  &   \\ \hline
$M_\rho$ [GeV] \platzoben  & 0.778 & 0.776  & \cite{pdg04} \\ \hline
$M_\omega$ [GeV] \platzoben  & 0.778 & 0.783  & \cite{pdg04} \\ \hline
$M_{K^*}$ [GeV] \platzoben  & 0.902 & 0.892 & \cite{pdg04}  \\ \hline
$M_\phi$ [GeV] \platzoben  & 1.012  & 1.019  & \cite{pdg04}  
\\ \hline \hline
$F_0$ [GeV] \platzoben  & 0.0971 & $0.086 \pm 0.010$  & \cite{gasleut2}  
\\ \hline
$d_F$ [1/GeV] \mehrplatzoben  & 0.0295 & --  &  \\ \hline
$F_{\rho\gamma}$   [GeV] \platzoben  & 0.138 & 0.154 & \cite{pdg04}  \\ \hline
$F_{\omega\gamma}$ [GeV] \platzoben  & 0.138 & 0.138 & \cite{pdg04}  \\ \hline
$F_{\phi\gamma}$ [GeV] \platzoben  & 0.162 & 0.162 & \cite{pdg04}  \\ \hline
$G_{\rho\pi\pi}$ [GeV] \platzoben  & 0.0692 & 0.0650 & \cite{pdg04} \\ \hline
$G_{\phi KK}$ [GeV] \platzoben  & 0.0808 & 0.0793 & \cite{pdg04}  
\\ \hline \hline
$L_1$ [$10^{-3}$]\platzoben  & 1.0 & $0.37\pm 0.23$  & \cite{Bijnens:1994ie}  
\\ \hline
$L_2$ [$10^{-3}$]\platzoben  & 2.0 & $1.35\pm 0.23$  & \cite{Bijnens:1994ie}  
\\ \hline
$L_5$ [$10^{-3}$]\platzoben  & 0.89 & $1.4\pm 0.5$  & \cite{gasleut2}  
\\ \hline
$F_\pi$ [GeV] \platzoben  & 0.0978 & 0.0924 & \cite{pdg04}  \\ \hline
$F_K$ [GeV] \platzoben  & 0.106 & 0.113 & \cite{pdg04}  \\ \hline
\end{tabular}
\caption{Properties of the vector meson nonet. The free parameters $F_0$ and
$d_F$ were fitted to $F_{\omega\gamma}$ and $F_{\phi\gamma}$. 
See main text for more details.}
\label{tab:vnonet}
\end{table}

As free parameters we take the pion decay constant in the chiral limit, $F_0$, and
the splitting parameter $d_F$ and fit them to the dilepton decay widths
of the narrow states $\omega$ and $\phi$. Results are shown in table \ref{tab:vnonet}.
Obviously, the results show an overall very satisfying agreement. 
In the following, we shall discuss these results in more detail.

Concerning the value of the pion decay constant in the chiral limit $F_0$ we observe
that it is on the upper end of the (rather large) allowed range.
Note that the chiral limit value obtained in \cite{gasleut1}
concerns the $SU(2)$ chiral limit where the up and down quark masses are sent to zero 
while the strange quark mass is kept fixed. Here we deal with the $SU(3)$ chiral limit
studied also in \cite{gasleut2}. 

From (\ref{eq:vnon-fvspl-rho}) it is obvious that $F_{\rho\gamma}$ and 
$F_{\omega\gamma}$ agree
in the large-$N_c$ limit. On the other hand, the experimental value of $F_{\rho\gamma}$ 
deviates from $F_{\omega\gamma}$ by about 10\%. 
One can take different points of view concerning this
deviation: In principle, one could be satisfied with a 10\% accuracy of a large-$N_c$
approach. On the other hand, we see that on average our fit is even much better. 
We recall that one can expect to find a better description of narrow resonances
within a large-$N_c$ treatment. 
In total, concerning the decay widths of the vector mesons, we observe that our
values for the coupling constants of the $\rho$-meson (which is the broadest resonance!)
show the largest deviations from experiment (about 10\%). All the other
coupling constants are reproduced extremely well.

Our value for $L_5$ is somewhat low. On the other
hand, the value for $L_5$ depends on the renormalization scale. This dependence is
subleading in $N_c$. From the practical point of view, however, the dependence is
rather large: The central value given in table \ref{tab:vnonet} varies from 2.4 to 0.8,
if the renormalization point changes from 0.5 to 1 GeV. 
More general, the values for the calculated $L_i$'s can easily change by a factor of 
two or more if the renormalization point changes from 0.5 to 1 GeV \cite{eckgas}.
Therefore, our somewhat large disagreement is not really significant.

Our values for the pion and kaon decay constants deviate less than 10\% from the
experimental value. We have obtained the correct sign for the splitting between
the decay constants (which is already not trivial). Quantitatively, however,
the splitting between $F_\pi$ and $F_K$ is underestimated. This can be traced back
to the too small value for $L_5$ which we have already discussed. Obviously, the physical
values for the pion and kaon decay constants are significantly influences by chiral
logs which are formally subleading in $1/N_c$. 

In general, we can be rather satisfied with our approach based on chiral symmetry
and the large-$N_c$ approximation. We close this section with some comments concerning
the comparison to some other approaches on the properties of vector mesons.
First of all, it is important to stress that the splitting
pattern of the coupling constant is experimentally significant: 
E.g.~the deviations between the 
experimental values of the different $G_i$'s are larger than 10\%. The same is true
for the different $F_i$'s. In approaches where $e_f^V$ and $e_g^V$ are neglected
(e.g.~\cite{Oller:2000ug}) these experimental differences must be generated by Goldstone
boson loops, i.e.~by subleading orders in $1/N_c$. In principle, we see no reason
why there should be no splitting in leading $N_c$ order. 

There is one case where a rather specific splitting pattern is introduced, namely
if the vector mesons are introduced as gauge bosons in one or the 
other way \cite{Meissner:1987ge,Bando:1987br,Harada:2003jx}. (Note that
this is not the approach used here.) In this case the following combinations
would all be equal to the coupling constant $g^2$ 
(cf.~(\ref{eq:grhopipi})):
\begin{equation}
  \label{eq:alleqno1}
{G_{\rho\pi\pi}^2 M_\rho^2 \over F_\pi^4} \,,
{G_{\phi KK}^2 M_\phi^2 \over F_K^4} \,,
\end{equation}
and the following combinations would all be equal to the coupling constant $g_\gamma^2$ 
(cf.~(\ref{eq:fvgcon})):
\begin{equation}
  \label{eq:alleqno2}
{M_\rho^2 \over F_{\rho\gamma}^2} \,,
{M_\omega^2 \over F_{\omega\gamma}^2} \,,
{M_\phi^2 \over F_{\phi\gamma}^2} \,.
\end{equation}
Note that these combinations appear as coefficients of the various decay widths
(\ref{eq:vn-dilepw}) and (\ref{eq:vn-hadrw}). The experimental values for these
coefficients are not really all equal as can be seen 
e.g.~in the tables of \cite{klingl1}. Our splitting pattern is parametrically 
very different from these gauge boson approaches. 
We conclude that our successful description of the splitting
pattern of the vector meson coupling constants casts some doubts on the introduction
of vector mesons as gauge bosons.
In principle, one might argue that a universal coupling constant is more economical
than our approach or, to argue the other way around, that it is not surprising that 
we get a better
description of the data since we have more free parameters. This, however, is not true:
Any approach has to introduce $F_0$ and $L_5$ as free parameters. Since our splitting
parameters for the coupling constants are related to each other and to $L_5$ and $F_0$
by (\ref{eq:resdfdgeg2}) we do not have more parameters than other approaches.

\section{Summary}
\label{sec:sum}

We have used the resonance saturation approach to determine properties
of the whole lowest-lying vector meson nonet. For that purpose the approach was
extended here beyond the chiral limit \cite{eckgas,Ecker:1989yg} not only for the
vector meson masses \cite{Cirigliano:2003yq} but also for their coupling constants.

From the conceptual point of view we have found that the correct assignment of 
the $N_c$-dependence of quark charges is mandatory to obtain a consistent picture
within a basically purely hadronic approach. For the future this might open the
possibility to determine the number of colors in a hadronic framework. In principle,
such a perspective does not come unexpected: In vector meson dominance approaches
with universal vector meson coupling constants the coupling of the $\rho$-meson
e.g.~to nucleons is suppressed by $1/N_c$ as compared to the $\omega$-meson
(see e.g.~\cite{Meissner:1987ge} and also the appendix of \cite{Leupold:2004gh}). 
This is just the factor $Q_u+Q_d = 2Y_q$ which appeared in (\ref{eq:vn-emcur}).

From the quantitative point of view the following was achieved in the present work:
With four input parameters we determined the vector meson nonet masses, 
the decay constants of pion and kaon, 
and the coupling constants for the decays of 
the neutral vector mesons into dileptons and two Goldstone bosons.
In general, we obtained very satisfying results within our approach.
In no case the deviation was larger than 10\%; usually it turned out to be much lower.

\acknowledgments The author acknowledges stimulating discussions with J.~R.~Pelaez.
He also thanks U.~Mosel for continuous support. Finally he thanks S.~Peris and
R.~Shrock for drawing his attention to \cite{Peris:1994dh} and \cite{Shrock:1995bp}, 
respectively.

\bibliography{literature}

\begin{thebibliography}{46}
\expandafter\ifx\csname natexlab\endcsname\relax\def\natexlab#1{#1}\fi
\expandafter\ifx\csname bibnamefont\endcsname\relax
  \def\bibnamefont#1{#1}\fi
\expandafter\ifx\csname bibfnamefont\endcsname\relax
  \def\bibfnamefont#1{#1}\fi
\expandafter\ifx\csname citenamefont\endcsname\relax
  \def\citenamefont#1{#1}\fi
\expandafter\ifx\csname url\endcsname\relax
  \def\url#1{\texttt{#1}}\fi
\expandafter\ifx\csname urlprefix\endcsname\relax\def\urlprefix{URL }\fi
\providecommand{\bibinfo}[2]{#2}
\providecommand{\eprint}[2][]{\url{#2}}

\bibitem[{\citenamefont{Gasser and Leutwyler}(1984)}]{gasleut1}
\bibinfo{author}{\bibfnamefont{J.}~\bibnamefont{Gasser}} \bibnamefont{and}
  \bibinfo{author}{\bibfnamefont{H.}~\bibnamefont{Leutwyler}},
  \bibinfo{journal}{Ann. Phys.} \textbf{\bibinfo{volume}{158}},
  \bibinfo{pages}{142} (\bibinfo{year}{1984}).

\bibitem[{\citenamefont{Gasser and Leutwyler}(1985)}]{gasleut2}
\bibinfo{author}{\bibfnamefont{J.}~\bibnamefont{Gasser}} \bibnamefont{and}
  \bibinfo{author}{\bibfnamefont{H.}~\bibnamefont{Leutwyler}},
  \bibinfo{journal}{Nucl.~Phys.} \textbf{\bibinfo{volume}{B250}},
  \bibinfo{pages}{465, 517, 539} (\bibinfo{year}{1985}).

\bibitem[{\citenamefont{Creutz}(1983)}]{Creutz:1984mg}
\bibinfo{author}{\bibfnamefont{M.}~\bibnamefont{Creutz}},
  \emph{\bibinfo{title}{Quarks, Gluons and Lattices}}, Cambridge Monographs On
  Mathematical Physics (\bibinfo{publisher}{Cambridge University Press},
  \bibinfo{address}{Cambridge, UK}, \bibinfo{year}{1983}).

\bibitem[{\citenamefont{Giusti et~al.}(2004)\citenamefont{Giusti, Hernandez,
  Laine, Weisz, and Wittig}}]{Giusti:2004yp}
\bibinfo{author}{\bibfnamefont{L.}~\bibnamefont{Giusti}},
  \bibinfo{author}{\bibfnamefont{P.}~\bibnamefont{Hernandez}},
  \bibinfo{author}{\bibfnamefont{M.}~\bibnamefont{Laine}},
  \bibinfo{author}{\bibfnamefont{P.}~\bibnamefont{Weisz}}, \bibnamefont{and}
  \bibinfo{author}{\bibfnamefont{H.}~\bibnamefont{Wittig}},
  \bibinfo{journal}{JHEP} \textbf{\bibinfo{volume}{04}}, \bibinfo{pages}{013}
  (\bibinfo{year}{2004}), \eprint{hep-lat/0402002}.

\bibitem[{\citenamefont{Amoros et~al.}(2001)\citenamefont{Amoros, Bijnens, and
  Talavera}}]{Amoros:2001cp}
\bibinfo{author}{\bibfnamefont{G.}~\bibnamefont{Amoros}},
  \bibinfo{author}{\bibfnamefont{J.}~\bibnamefont{Bijnens}}, \bibnamefont{and}
  \bibinfo{author}{\bibfnamefont{P.}~\bibnamefont{Talavera}},
  \bibinfo{journal}{Nucl. Phys.} \textbf{\bibinfo{volume}{B602}},
  \bibinfo{pages}{87} (\bibinfo{year}{2001}), \eprint{hep-ph/0101127}.

\bibitem[{\citenamefont{Ecker et~al.}(1989{\natexlab{a}})\citenamefont{Ecker,
  Gasser, Pich, and de~Rafael}}]{eckgas}
\bibinfo{author}{\bibfnamefont{G.}~\bibnamefont{Ecker}},
  \bibinfo{author}{\bibfnamefont{J.}~\bibnamefont{Gasser}},
  \bibinfo{author}{\bibfnamefont{A.}~\bibnamefont{Pich}}, \bibnamefont{and}
  \bibinfo{author}{\bibfnamefont{E.}~\bibnamefont{de~Rafael}},
  \bibinfo{journal}{Nucl. Phys.} \textbf{\bibinfo{volume}{B321}},
  \bibinfo{pages}{311} (\bibinfo{year}{1989}{\natexlab{a}}).

\bibitem[{\citenamefont{Ecker et~al.}(1989{\natexlab{b}})\citenamefont{Ecker,
  Gasser, Leutwyler, Pich, and de~Rafael}}]{Ecker:1989yg}
\bibinfo{author}{\bibfnamefont{G.}~\bibnamefont{Ecker}},
  \bibinfo{author}{\bibfnamefont{J.}~\bibnamefont{Gasser}},
  \bibinfo{author}{\bibfnamefont{H.}~\bibnamefont{Leutwyler}},
  \bibinfo{author}{\bibfnamefont{A.}~\bibnamefont{Pich}}, \bibnamefont{and}
  \bibinfo{author}{\bibfnamefont{E.}~\bibnamefont{de~Rafael}},
  \bibinfo{journal}{Phys. Lett.} \textbf{\bibinfo{volume}{B223}},
  \bibinfo{pages}{425} (\bibinfo{year}{1989}{\natexlab{b}}).

\bibitem[{\citenamefont{Donoghue et~al.}(1989)\citenamefont{Donoghue, Ramirez,
  and Valencia}}]{Donoghue:1989ed}
\bibinfo{author}{\bibfnamefont{J.~F.} \bibnamefont{Donoghue}},
  \bibinfo{author}{\bibfnamefont{C.}~\bibnamefont{Ramirez}}, \bibnamefont{and}
  \bibinfo{author}{\bibfnamefont{G.}~\bibnamefont{Valencia}},
  \bibinfo{journal}{Phys. Rev.} \textbf{\bibinfo{volume}{D39}},
  \bibinfo{pages}{1947} (\bibinfo{year}{1989}).

\bibitem[{\citenamefont{Gomez~Nicola and Pelaez}(2002)}]{pelaez02}
\bibinfo{author}{\bibfnamefont{A.}~\bibnamefont{Gomez~Nicola}}
  \bibnamefont{and} \bibinfo{author}{\bibfnamefont{J.~R.}
  \bibnamefont{Pelaez}}, \bibinfo{journal}{Phys. Rev.}
  \textbf{\bibinfo{volume}{D65}}, \bibinfo{pages}{054009}
  (\bibinfo{year}{2002}), \eprint{hep-ph/0109056}.

\bibitem[{\citenamefont{Diakonov and Petrov}(1986)}]{diakpet}
\bibinfo{author}{\bibfnamefont{D.}~\bibnamefont{Diakonov}} \bibnamefont{and}
  \bibinfo{author}{\bibfnamefont{V.~Y.} \bibnamefont{Petrov}},
  \bibinfo{journal}{Nucl. Phys.} \textbf{\bibinfo{volume}{B272}},
  \bibinfo{pages}{457} (\bibinfo{year}{1986}).

\bibitem[{\citenamefont{Espriu et~al.}(1990)\citenamefont{Espriu, de~Rafael,
  and Taron}}]{espraf}
\bibinfo{author}{\bibfnamefont{D.}~\bibnamefont{Espriu}},
  \bibinfo{author}{\bibfnamefont{E.}~\bibnamefont{de~Rafael}},
  \bibnamefont{and} \bibinfo{author}{\bibfnamefont{J.}~\bibnamefont{Taron}},
  \bibinfo{journal}{Nucl. Phys.} \textbf{\bibinfo{volume}{B345}},
  \bibinfo{pages}{22} (\bibinfo{year}{1990}), \bibinfo{note}{erratum-ibid. {\bf
  B355}, 278 (1991)}.

\bibitem[{\citenamefont{Sch\"uren et~al.}(1992)\citenamefont{Sch\"uren,
  Ruiz~Arriola, and Goeke}}]{Schuren:1992sc}
\bibinfo{author}{\bibfnamefont{C.}~\bibnamefont{Sch\"uren}},
  \bibinfo{author}{\bibfnamefont{E.}~\bibnamefont{Ruiz~Arriola}},
  \bibnamefont{and} \bibinfo{author}{\bibfnamefont{K.}~\bibnamefont{Goeke}},
  \bibinfo{journal}{Nucl. Phys.} \textbf{\bibinfo{volume}{A547}},
  \bibinfo{pages}{612} (\bibinfo{year}{1992}).

\bibitem[{\citenamefont{M\"uller and Klevansky}(1994)}]{Mueller:1994dh}
\bibinfo{author}{\bibfnamefont{J.}~\bibnamefont{M\"uller}} \bibnamefont{and}
  \bibinfo{author}{\bibfnamefont{S.~P.} \bibnamefont{Klevansky}},
  \bibinfo{journal}{Phys. Rev.} \textbf{\bibinfo{volume}{C50}},
  \bibinfo{pages}{410} (\bibinfo{year}{1994}).

\bibitem[{\citenamefont{Pich}(1995)}]{Pich:1995bw}
\bibinfo{author}{\bibfnamefont{A.}~\bibnamefont{Pich}}, \bibinfo{journal}{Rept.
  Prog. Phys.} \textbf{\bibinfo{volume}{58}}, \bibinfo{pages}{563}
  (\bibinfo{year}{1995}), \eprint{hep-ph/9502366}.

\bibitem[{\citenamefont{Leupold}(2004)}]{Leupold:2003zb}
\bibinfo{author}{\bibfnamefont{S.}~\bibnamefont{Leupold}},
  \bibinfo{journal}{Nucl. Phys.} \textbf{\bibinfo{volume}{A743}},
  \bibinfo{pages}{283} (\bibinfo{year}{2004}), \eprint{hep-ph/0303020}.

\bibitem[{\citenamefont{Peris et~al.}(1998)\citenamefont{Peris, Perrottet, and
  de~Rafael}}]{Peris:1998nj}
\bibinfo{author}{\bibfnamefont{S.}~\bibnamefont{Peris}},
  \bibinfo{author}{\bibfnamefont{M.}~\bibnamefont{Perrottet}},
  \bibnamefont{and}
  \bibinfo{author}{\bibfnamefont{E.}~\bibnamefont{de~Rafael}},
  \bibinfo{journal}{JHEP} \textbf{\bibinfo{volume}{05}}, \bibinfo{pages}{011}
  (\bibinfo{year}{1998}), \eprint{hep-ph/9805442}.

\bibitem[{\citenamefont{Kaiser and Leutwyler}(2000)}]{Kaiser:2000gs}
\bibinfo{author}{\bibfnamefont{R.}~\bibnamefont{Kaiser}} \bibnamefont{and}
  \bibinfo{author}{\bibfnamefont{H.}~\bibnamefont{Leutwyler}},
  \bibinfo{journal}{Eur. Phys. J.} \textbf{\bibinfo{volume}{C17}},
  \bibinfo{pages}{623} (\bibinfo{year}{2000}), \eprint{hep-ph/0007101}.

\bibitem[{\citenamefont{Peris and de~Rafael}(1995)}]{Peris:1994dh}
\bibinfo{author}{\bibfnamefont{S.}~\bibnamefont{Peris}} \bibnamefont{and}
  \bibinfo{author}{\bibfnamefont{E.}~\bibnamefont{de~Rafael}},
  \bibinfo{journal}{Phys. Lett.} \textbf{\bibinfo{volume}{B348}},
  \bibinfo{pages}{539} (\bibinfo{year}{1995}), \eprint{hep-ph/9412343}.

\bibitem[{\citenamefont{Ecker}(1995)}]{Ecker:1995gg}
\bibinfo{author}{\bibfnamefont{G.}~\bibnamefont{Ecker}},
  \bibinfo{journal}{Prog. Part. Nucl. Phys.} \textbf{\bibinfo{volume}{35}},
  \bibinfo{pages}{1} (\bibinfo{year}{1995}), \eprint{hep-ph/9501357}.

\bibitem[{\citenamefont{Scherer}(2003)}]{Scherer:2002tk}
\bibinfo{author}{\bibfnamefont{S.}~\bibnamefont{Scherer}},
  \bibinfo{journal}{Adv. Nucl. Phys.} \textbf{\bibinfo{volume}{27}},
  \bibinfo{pages}{277} (\bibinfo{year}{2003}), \eprint{hep-ph/0210398}.

\bibitem[{\citenamefont{Cirigliano et~al.}(2003)\citenamefont{Cirigliano,
  Ecker, Neufeld, and Pich}}]{Cirigliano:2003yq}
\bibinfo{author}{\bibfnamefont{V.}~\bibnamefont{Cirigliano}},
  \bibinfo{author}{\bibfnamefont{G.}~\bibnamefont{Ecker}},
  \bibinfo{author}{\bibfnamefont{H.}~\bibnamefont{Neufeld}}, \bibnamefont{and}
  \bibinfo{author}{\bibfnamefont{A.}~\bibnamefont{Pich}},
  \bibinfo{journal}{JHEP} \textbf{\bibinfo{volume}{06}}, \bibinfo{pages}{012}
  (\bibinfo{year}{2003}), \eprint{hep-ph/0305311}.

\bibitem[{\citenamefont{Rosell et~al.}(2004)\citenamefont{Rosell, Sanz-Cillero,
  and Pich}}]{Rosell:2004mn}
\bibinfo{author}{\bibfnamefont{I.}~\bibnamefont{Rosell}},
  \bibinfo{author}{\bibfnamefont{J.~J.} \bibnamefont{Sanz-Cillero}},
  \bibnamefont{and} \bibinfo{author}{\bibfnamefont{A.}~\bibnamefont{Pich}},
  \bibinfo{journal}{JHEP} \textbf{\bibinfo{volume}{08}}, \bibinfo{pages}{042}
  (\bibinfo{year}{2004}), \eprint{hep-ph/0407240}.

\bibitem[{\citenamefont{B\"ar and Wiese}(2001)}]{Bar:2001qk}
\bibinfo{author}{\bibfnamefont{O.}~\bibnamefont{B\"ar}} \bibnamefont{and}
  \bibinfo{author}{\bibfnamefont{U.~J.} \bibnamefont{Wiese}},
  \bibinfo{journal}{Nucl. Phys.} \textbf{\bibinfo{volume}{B609}},
  \bibinfo{pages}{225} (\bibinfo{year}{2001}), \eprint{hep-ph/0105258}.

\bibitem[{\citenamefont{Abbas}(1990)}]{Abbas:1990kd}
\bibinfo{author}{\bibfnamefont{A.}~\bibnamefont{Abbas}},
  \bibinfo{journal}{Phys. Lett.} \textbf{\bibinfo{volume}{B238}},
  \bibinfo{pages}{344} (\bibinfo{year}{1990}).

\bibitem[{\citenamefont{Shrock}(1996)}]{Shrock:1995bp}
\bibinfo{author}{\bibfnamefont{R.}~\bibnamefont{Shrock}},
  \bibinfo{journal}{Phys. Rev.} \textbf{\bibinfo{volume}{D53}},
  \bibinfo{pages}{6465} (\bibinfo{year}{1996}), \eprint{hep-ph/9512430}.

\bibitem[{\citenamefont{Abbas}(2000)}]{Abbas:2000xk}
\bibinfo{author}{\bibfnamefont{A.}~\bibnamefont{Abbas}} (\bibinfo{year}{2000}),
  \eprint{hep-ph/0009242}.

\bibitem[{\citenamefont{'t~Hooft}(1974)}]{'tHooft:1974jz}
\bibinfo{author}{\bibfnamefont{G.}~\bibnamefont{'t~Hooft}},
  \bibinfo{journal}{Nucl. Phys.} \textbf{\bibinfo{volume}{B72}},
  \bibinfo{pages}{461} (\bibinfo{year}{1974}).

\bibitem[{\citenamefont{Eidelman et~al.}(2004)}]{pdg04}
\bibinfo{author}{\bibfnamefont{S.}~\bibnamefont{Eidelman}} \bibnamefont{et~al.}
  (\bibinfo{collaboration}{Particle Data Group}), \bibinfo{journal}{Phys.
  Lett.} \textbf{\bibinfo{volume}{B592}}, \bibinfo{pages}{1}
  (\bibinfo{year}{2004}).

\bibitem[{\citenamefont{Klingl et~al.}(1996)\citenamefont{Klingl, Kaiser, and
  Weise}}]{klingl1}
\bibinfo{author}{\bibfnamefont{F.}~\bibnamefont{Klingl}},
  \bibinfo{author}{\bibfnamefont{N.}~\bibnamefont{Kaiser}}, \bibnamefont{and}
  \bibinfo{author}{\bibfnamefont{W.}~\bibnamefont{Weise}}, \bibinfo{journal}{Z.
  Phys.} \textbf{\bibinfo{volume}{A356}}, \bibinfo{pages}{193}
  (\bibinfo{year}{1996}), \eprint{hep-ph/9607431}.

\bibitem[{\citenamefont{Kawarabayashi and Suzuki}(1966)}]{ks}
\bibinfo{author}{\bibfnamefont{K.}~\bibnamefont{Kawarabayashi}}
  \bibnamefont{and} \bibinfo{author}{\bibfnamefont{M.}~\bibnamefont{Suzuki}},
  \bibinfo{journal}{Phys. Rev. Lett.} \textbf{\bibinfo{volume}{16}},
  \bibinfo{pages}{255} (\bibinfo{year}{1966}).

\bibitem[{\citenamefont{Riazuddin and Fayyazuddin}(1966)}]{fr}
\bibinfo{author}{\bibnamefont{Riazuddin}} \bibnamefont{and}
  \bibinfo{author}{\bibnamefont{Fayyazuddin}}, \bibinfo{journal}{Phys. Rev.}
  \textbf{\bibinfo{volume}{147}}, \bibinfo{pages}{1071} (\bibinfo{year}{1966}).

\bibitem[{\citenamefont{Herrmann et~al.}(1993)\citenamefont{Herrmann, Friman,
  and N\"orenberg}}]{herrmann}
\bibinfo{author}{\bibfnamefont{M.}~\bibnamefont{Herrmann}},
  \bibinfo{author}{\bibfnamefont{B.~L.} \bibnamefont{Friman}},
  \bibnamefont{and}
  \bibinfo{author}{\bibfnamefont{W.}~\bibnamefont{N\"orenberg}},
  \bibinfo{journal}{Nucl. Phys.} \textbf{\bibinfo{volume}{A560}},
  \bibinfo{pages}{411} (\bibinfo{year}{1993}).

\bibitem[{\citenamefont{Dung and Truong}(1996)}]{Dung:1996rp}
\bibinfo{author}{\bibfnamefont{L.~v.} \bibnamefont{Dung}} \bibnamefont{and}
  \bibinfo{author}{\bibfnamefont{T.~N.} \bibnamefont{Truong}}
  (\bibinfo{year}{1996}), \eprint{hep-ph/9607378}.

\bibitem[{\citenamefont{Djukanovic et~al.}(2004)\citenamefont{Djukanovic,
  Schindler, Gegelia, Japaridze, and Scherer}}]{Djukanovic:2004mm}
\bibinfo{author}{\bibfnamefont{D.}~\bibnamefont{Djukanovic}},
  \bibinfo{author}{\bibfnamefont{M.~R.} \bibnamefont{Schindler}},
  \bibinfo{author}{\bibfnamefont{J.}~\bibnamefont{Gegelia}},
  \bibinfo{author}{\bibfnamefont{G.}~\bibnamefont{Japaridze}},
  \bibnamefont{and} \bibinfo{author}{\bibfnamefont{S.}~\bibnamefont{Scherer}},
  \bibinfo{journal}{Phys. Rev. Lett.} \textbf{\bibinfo{volume}{93}},
  \bibinfo{pages}{122002} (\bibinfo{year}{2004}), \eprint{hep-ph/0407239}.

\bibitem[{\citenamefont{Djukanovic et~al.}(2005)\citenamefont{Djukanovic,
  Schindler, Gegelia, and Scherer}}]{Djukanovic:2005ag}
\bibinfo{author}{\bibfnamefont{D.}~\bibnamefont{Djukanovic}},
  \bibinfo{author}{\bibfnamefont{M.~R.} \bibnamefont{Schindler}},
  \bibinfo{author}{\bibfnamefont{J.}~\bibnamefont{Gegelia}}, \bibnamefont{and}
  \bibinfo{author}{\bibfnamefont{S.}~\bibnamefont{Scherer}}
  (\bibinfo{year}{2005}), \eprint{hep-ph/0505180}.

\bibitem[{\citenamefont{Gell-Mann et~al.}(1968)\citenamefont{Gell-Mann, Oakes,
  and Renner}}]{GOR}
\bibinfo{author}{\bibfnamefont{M.}~\bibnamefont{Gell-Mann}},
  \bibinfo{author}{\bibfnamefont{R.~J.} \bibnamefont{Oakes}}, \bibnamefont{and}
  \bibinfo{author}{\bibfnamefont{B.}~\bibnamefont{Renner}},
  \bibinfo{journal}{Phys. Rev.} \textbf{\bibinfo{volume}{175}},
  \bibinfo{pages}{2195} (\bibinfo{year}{1968}).

\bibitem[{\citenamefont{Borasoy}(2004)}]{Borasoy:2004ua}
\bibinfo{author}{\bibfnamefont{B.}~\bibnamefont{Borasoy}},
  \bibinfo{journal}{Eur. Phys. J.} \textbf{\bibinfo{volume}{C34}},
  \bibinfo{pages}{317} (\bibinfo{year}{2004}), \eprint{hep-ph/0402294}.

\bibitem[{\citenamefont{Borasoy and Lipartia}(2005)}]{Borasoy:2004mf}
\bibinfo{author}{\bibfnamefont{B.}~\bibnamefont{Borasoy}} \bibnamefont{and}
  \bibinfo{author}{\bibfnamefont{E.}~\bibnamefont{Lipartia}},
  \bibinfo{journal}{Phys. Rev.} \textbf{\bibinfo{volume}{D71}},
  \bibinfo{pages}{014027} (\bibinfo{year}{2005}), \eprint{hep-ph/0410141}.

\bibitem[{\citenamefont{Shifman et~al.}(1979)\citenamefont{Shifman, Vainshtein,
  and Zakharov}}]{shif79}
\bibinfo{author}{\bibfnamefont{M.~A.} \bibnamefont{Shifman}},
  \bibinfo{author}{\bibfnamefont{A.~I.} \bibnamefont{Vainshtein}},
  \bibnamefont{and} \bibinfo{author}{\bibfnamefont{V.~I.}
  \bibnamefont{Zakharov}}, \bibinfo{journal}{Nucl.~Phys.}
  \textbf{\bibinfo{volume}{B147}}, \bibinfo{pages}{385, 448}
  (\bibinfo{year}{1979}).

\bibitem[{\citenamefont{Reinders et~al.}(1985)\citenamefont{Reinders,
  Rubinstein, and Yazaki}}]{Reinders:1984sr}
\bibinfo{author}{\bibfnamefont{L.~J.} \bibnamefont{Reinders}},
  \bibinfo{author}{\bibfnamefont{H.}~\bibnamefont{Rubinstein}},
  \bibnamefont{and} \bibinfo{author}{\bibfnamefont{S.}~\bibnamefont{Yazaki}},
  \bibinfo{journal}{Phys. Rept.} \textbf{\bibinfo{volume}{127}},
  \bibinfo{pages}{1} (\bibinfo{year}{1985}).

\bibitem[{\citenamefont{Bijnens et~al.}(1994)\citenamefont{Bijnens, Colangelo,
  and Gasser}}]{Bijnens:1994ie}
\bibinfo{author}{\bibfnamefont{J.}~\bibnamefont{Bijnens}},
  \bibinfo{author}{\bibfnamefont{G.}~\bibnamefont{Colangelo}},
  \bibnamefont{and} \bibinfo{author}{\bibfnamefont{J.}~\bibnamefont{Gasser}},
  \bibinfo{journal}{Nucl. Phys.} \textbf{\bibinfo{volume}{B427}},
  \bibinfo{pages}{427} (\bibinfo{year}{1994}), \eprint{hep-ph/9403390}.

\bibitem[{\citenamefont{Oller et~al.}(2001)\citenamefont{Oller, Oset, and
  Palomar}}]{Oller:2000ug}
\bibinfo{author}{\bibfnamefont{J.~A.} \bibnamefont{Oller}},
  \bibinfo{author}{\bibfnamefont{E.}~\bibnamefont{Oset}}, \bibnamefont{and}
  \bibinfo{author}{\bibfnamefont{J.~E.} \bibnamefont{Palomar}},
  \bibinfo{journal}{Phys. Rev.} \textbf{\bibinfo{volume}{D63}},
  \bibinfo{pages}{114009} (\bibinfo{year}{2001}), \eprint{hep-ph/0011096}.

\bibitem[{\citenamefont{Meissner}(1988)}]{Meissner:1987ge}
\bibinfo{author}{\bibfnamefont{U.~G.} \bibnamefont{Meissner}},
  \bibinfo{journal}{Phys. Rept.} \textbf{\bibinfo{volume}{161}},
  \bibinfo{pages}{213} (\bibinfo{year}{1988}).

\bibitem[{\citenamefont{Bando et~al.}(1988)\citenamefont{Bando, Kugo, and
  Yamawaki}}]{Bando:1987br}
\bibinfo{author}{\bibfnamefont{M.}~\bibnamefont{Bando}},
  \bibinfo{author}{\bibfnamefont{T.}~\bibnamefont{Kugo}}, \bibnamefont{and}
  \bibinfo{author}{\bibfnamefont{K.}~\bibnamefont{Yamawaki}},
  \bibinfo{journal}{Phys. Rept.} \textbf{\bibinfo{volume}{164}},
  \bibinfo{pages}{217} (\bibinfo{year}{1988}).

\bibitem[{\citenamefont{Harada and Yamawaki}(2003)}]{Harada:2003jx}
\bibinfo{author}{\bibfnamefont{M.}~\bibnamefont{Harada}} \bibnamefont{and}
  \bibinfo{author}{\bibfnamefont{K.}~\bibnamefont{Yamawaki}},
  \bibinfo{journal}{Phys. Rept.} \textbf{\bibinfo{volume}{381}},
  \bibinfo{pages}{1} (\bibinfo{year}{2003}), \eprint{hep-ph/0302103}.

\bibitem[{\citenamefont{Leupold and Post}(2005)}]{Leupold:2004gh}
\bibinfo{author}{\bibfnamefont{S.}~\bibnamefont{Leupold}} \bibnamefont{and}
  \bibinfo{author}{\bibfnamefont{M.}~\bibnamefont{Post}},
  \bibinfo{journal}{Nucl. Phys.} \textbf{\bibinfo{volume}{A747}},
  \bibinfo{pages}{425} (\bibinfo{year}{2005}), \eprint{nucl-th/0402048}.

\end{thebibliography}
\bibliographystyle{apsrev}

\end{document}